\documentclass{ws-ijmpe}
\usepackage{color}

\def\be{\begin{equation}}
\def\ee{\end{equation}}
\def\bea{\begin{eqnarray}}
\def\eea{\end{eqnarray}}
\def\bfr{{\bf r}}
\def\bfq{{\bf q}}
\def\bfp{{\bf p}}

\def\fig#1{Fig.~\ref{#1}}
\def\eq#1{(\ref{#1})}
\def\papa#1#2{\frac{\partial#1}{\partial#2}}
\def\Cosh{{\rm cosh}}
\def\Sinh{{\rm sinh}}
\def\Tau{{\cal T}}
\def\lambdab{\widetilde{\lambda}}
\def\d{{\rm d}}

\begin{document}

\markboth{Brack and Roccia}{Closed-orbit theory for spatial
                            density oscillations}

%
\catchline{}{}{}{}{}
%

\title{CLOSED-ORBIT THEORY FOR SPATIAL DENSITY OSCILLATIONS}

\author{\footnotesize MATTHIAS BRACK and J\'ER\^OME ROCCIA} 

\address{Institute of Theoretical Physics, University of Regensburg,
D-93040 Regensburg, Germany\\
e-mail: matthias.brack@physik.uni-regensburg.de}

\maketitle

\begin{history}
\received{(\today)}
\revised{(revised date)}
\end{history}

\begin{abstract}
We briefly review a recently developed semiclassical theory\cite{rb} 
for quantum oscillations in the spatial (particle and kinetic energy) 
densities of finite fermion systems and present some examples of its 
results. We then discuss the inclusion of correlations (finite 
temperatures, pairing correlations) in the semiclassical theory.
\end{abstract}

\section{Introduction}
\label{secint}

We have recently proposed\cite{rb} a semiclassical theory for
quantum oscillations in the local particle
densities and kinetic-energy densities of a system of $N$ fermions 
in a local potential in $D$ dimensions, described by the stationary 
Schr\"odinger equation
\be
\left\{-{\hbar^2\over 2m} \nabla^2 + V(\bfr)\right\} \phi_n(\bfr) 
= E_n\, \phi_n(\bfr)\,.
\label{seq}
\ee
The potential 
$V(\bfr)$ can be considered to represent the self-consistent mean field 
of an interacting system of fermions obtained in density functional 
theory (DFT).\cite{dft} The single-particle wavefunctions 
$\phi_n(\bfr)$ then are the Kohn-Sham orbitals\cite{ks} and 
$\rho(\bfr)$ in \eq{rho} is the (ideally exact) ground-state 
particle density of the interacting system.\cite{hk}

Ordering the spectrum $\{E_n\}$ and choosing the energy scale such 
that $0 < E_1 \leq E_2 \leq \dots \leq E_n \leq \dots$, we fill the 
lowest levels up to the Fermi energy $\lambda$ and define the 
particle density by
\be
\rho(\bfr) \; := \; 2\!\! \sum_{E_n\leq \lambda}\!\! |\phi_n(\bfr)|^2, 
\qquad \int \rho(\bfr)\,\d^Dr = N\,.
\label{rho}
\ee
The factor 2 accounts for the spin degeneracy (the number $N$ is assumed 
to be even). 
Further degeneracies, which may arise for systems in $D>1$ dimensions,
will not be spelled out bout included in the summations over $n$.
For the kinetic-energy density, we consider two different definitions
\be
\tau(\bfr)   \; := \; - \frac{\hbar^2}{2m}\; 2\!\! \sum_{E_n\leq \lambda}
                     \!\! \phi_n^*(\bfr)\nabla^2 \phi_n(\bfr)\,,\qquad
\tau_1(\bfr) \; := \; \frac{\hbar^2}{2m}\; 2\!\! \sum_{E_n\leq \lambda}
                   \!\! |\nabla\phi_n(\bfr)|^2,
\label{tau}
\ee
which upon integration both yield the exact total kinetic energy.

\newpage

The density of states $g(E)$ of the system \eq{seq} is given by
\be
g(E) = \sum_n \delta(E-E_n)\,, \qquad N = N(\lambda) = 2\!\int_0^\lambda \d E\,g(E)\,.
\label{dos}
\ee
Separating its smooth and oscillatory parts by defining
\be
g(E):={\widetilde g}(E)+\delta g(E)\,,
\label{dossep}
\ee
the smooth part ${\widetilde g}(E)$ is given by the extended
Thomas-Fermi (ETF) theory (see chapter 4.4.3 of Ref.\cite{book}), 
while the the oscillating part $\delta g(E)$ can be described, to 
leading order in $\hbar$, by the {\it semiclassical trace 
formula}\cite{gutz,gubu}
\be
\delta g(E) \simeq \sum_{\text{PO}} {\cal A}_{\text{PO}}(E) \cos\left[
                   \frac{1}{\hbar}\,S_{\text{PO}}(E)-\frac{\pi}{2}\,\sigma_{\text{PO}}\right]. 
\label{trf}
\ee
The sum here is over all {\it periodic orbits} (POs) of the corresponding
classical system described by the Hamilton function $H(\bfq,\bfp)=\bfp^2\!/2m+V(\bfq)$. 
$S_{\text{PO}}(E)$ is the action integral along the periodic orbit:
\be
S_{\text{PO}}(E) = \oint_{\text{PO}} \bfp(E,{\bf q})\cdot \d {\bf q}\,,
\label{spo}
\ee
with the classical momentum given by $\bfp(E,\bfr)=
(\dot{{\bf r}}/|{\dot{\bf r}}|)\sqrt{2m[E-V(\bfr)]}$.
For systems in which all orbits are isolated in phase space, explicit 
expressions for the amplitudes ${\cal A}_{\text{PO}}(E)$, which depend on the 
stabilities of the orbits, and for the Maslov indices $\sigma_{\text{PO}}$
have been given by Gutzwiller.\cite{gutz} For systems with continuous 
symmetries and for integrable systems, alternative expressions for the 
amplitudes and Maslov indices have been derived by many authors; they 
may be found in Ref.\cite{book}

Separating smooth and oscillating terms of the spatial densities
\be
\rho(\bfr) := {\widetilde\rho}(\bfr)+\delta\rho(\bfr)\,,\quad
\tau(\bfr) := {\widetilde\tau}(\bfr)+\delta\tau(\bfr)\,,\quad
\tau_1(\bfr) := {\widetilde\tau}_1(\bfr)+\delta\tau_1(\bfr)\,,
\ee
the smooth parts are given by the ETF theory. For their 
oscillating parts we have obtained\cite{rb} the following semiclassical 
expressions, valid again to leading order in $\hbar$:
\begin{eqnarray}
\delta \rho(\bfr) & \simeq & \sum_\gamma {\cal A}_\gamma(\lambdab,\bfr)
                             \,\cos\left[\Phi_\gamma(\lambdab,\bfr)\right],
\label{drhosc}\\
\delta \tau(\bfr) & \simeq & \frac{\bfp^2(\lambdab,\bfr)}{2m}\,\sum_\gamma 
                             {\cal A}_\gamma(\lambdab,\bfr)
                             \,\cos\left[\Phi_\gamma(\lambdab,\bfr)\right],
\label{dtausc}\\
\delta\tau_1(\bfr)& \simeq & \frac{\bfp^2(\lambdab,\bfr)}{2m}\,\sum_\gamma 
                             {\cal A}_\gamma(\lambdab,\bfr)\,Q_\gamma(\lambdab,\bfr)
                             \,\cos\left[\Phi_\gamma(\lambdab,\bfr)\right].
\label{dtau1sc}
\end{eqnarray}
The sum here is over all {\it closed orbits} $\gamma$ starting and
ending in the point $\bfr$, and
\be
\Phi_\gamma(\lambdab,\bfr) = \frac{1}{\hbar}S_\gamma(\lambdab,\bfr)
                             -\frac{\pi}{2}\,\mu_\gamma-\frac{\pi}{4}\,(D+1)\,.
\label{phaser}
\ee
The action function $S_\gamma(\lambdab,\bfr)=S_\gamma(\lambdab,\bfr,\bfr'=\bfr)$ 
is gained from the general open action integral for an orbit starting
at $\bfr$ and ending at $\bfr'$ at fixed energy $E=\lambdab$


\be
S_\gamma(\lambdab,\bfr,\bfr') = \int_{\bfr}^{\bfr'} {\bf p}(\lambdab,{\bf q})
                                \cdot \d\,{\bf q}\,,
\label{actint}
\ee
and $\mu_\gamma$ is the Morse index that counts the number of conjugate points 
along the orbit.\cite{gutz,gubu} For the functions ${\cal A}_\gamma(\lambdab,\bfr)$ and 
$Q_\gamma(\lambdab,\bfr)$ we refer to our articles.\cite{rb,rbkm,br} The quantity
$\lambdab$ is the Fermi energy of the smooth (ETF) system, defined by
\be
\lambda = \lambdab + \delta\lambda\,, \qquad 
N = 2\int_0^{\lambdab} \d E\,{\widetilde g}(E)\,.
\ee
Since for POs the action integral $S_{\text{PO}}(\lambdab)$ is independent of 
$\bfr$, they do not yield any oscillating phases in the above expressions;
their contributions vary only smoothly with $\bfr$ through 
${\cal A}_{\rm PO}(\lambdab,\bfr)$ and $Q_{\rm  PO}(\lambdab,\bfr)$. The leading 
contributions to the {\it density oscillations} come from the {\it 
non-periodic orbits} (NPOs). For one-dimensional systems ($D$=1) it has, 
in fact, been shown\cite{rb,rbkm} that the contributions of the POs are 
completely absorbed by the smooth (TF) densities. In higher-dimensional 
systems, the POs must be included in \eq{drhosc} - \eq{dtau1sc} in
connection with symmetry breaking at $\bfr=0$ for spherical systems, 
and with bifurcations at finite distances $|\bfr|>0$ in general, as 
demonstrated explicitly for the two-dimensional circular billiard.\cite{br} 

\section{Selected results}
\label{secres}

In this section we give some selected results of our semiclassical theory. 
We first present a very general result that may have interesting 
consequences for DFT. From \eq{drhosc}, \eq{dtausc} one finds directly 
-- without knowledge of the orbits $\gamma$ -- the relation
\begin{equation}
\delta \tau({\bf r})\simeq[\lambdab-V({\bf r})] \,\delta \rho({\bf r}) \,,
\label{lvt}
\end{equation}
which we call the (differential) {\it local virial theorem} (LVT)
because it relates the potential and kinetic-energy densities
{\it locally} at any given point $\bfr$. The relation \eq{lvt} was 
derived\cite{bm} for isotropic harmonic oscillators in arbitrary
dimensions from their quantum-mechanical densities in the asymptotic
limit $N\to\infty$. In our semiclassical theory it is obtained 
for arbitrary potentials. Since no assumption 
about the potential or the nature of the closed orbits $\gamma$ must 
be made to derive the LVT \eq{lvt}, it holds for {\it arbitrary} 
(integrable or non-integrable) systems in arbitrary dimensions with 
a local potential $V(\bfr)$, and 
hence also for interacting fermions in the mean-field approximation 
given by the DFT. We recall, however, that \eq{lvt} is not expected 
to be valid close to the classical turning points where the
semiclassical expressions \eq{drhosc} - \eq{dtau1sc} diverge and must
be regularized by appropriate uniform approximations.\cite{rbkm,br}

A direct consequence of the LVT in \eq{lvt} is the following relation:
\be
\tau(\bfr) \; \simeq \; \tau_{\text{TF}}[\rho(\bfr)]\,.
\label{tfofrho}
\ee
Hereby
$\tau_{\text{TF}}[\rho_{\text{TF}}(\bfr)]=\tau_{\text{TF}}(\bfr)$ is the exact 
functional relation between the TF kinetic-energy and particle densities. 
Eq.\ \eq{tfofrho} states that this TF functional (without gradient
corrections!) holds approximately, 
for arbitrary local potentials $V(\bfr)$, also between the {\it exact 
quantum-mechanical densities $\tau(\bfr)$ and $\rho(\bfr)$} including 
their quantum oscillations. [It was shown in Ref.\cite{rbkm} to be
exact up to first order in $\delta\rho(\bfr)$.]

\begin{figure}[h]
\begin{center} 
\includegraphics[width=0.6\columnwidth,clip=true]{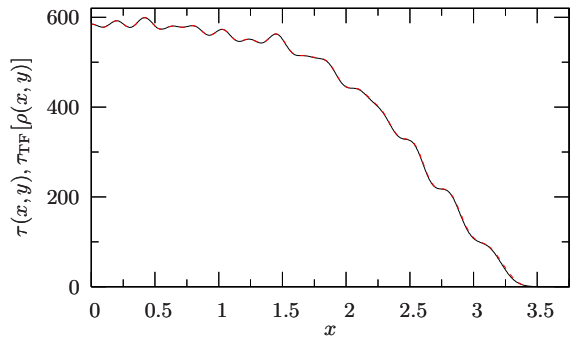}\vspace*{-0.3cm}
\caption{
TF relation \eq{tfofrho} for the potential
\eq{vqo} with $\kappa=0.6$ (units: $\hbar=m=1$) with $N=632$. 
Cuts along the diagonal $x=y$. The solid line is the l.h.s., and 
the dashed line is the r.h.s.\ of \eq{tfofrho}.}\label{chaos} 
\end{center}
\end{figure}

In \fig{chaos} we test \eq{tfofrho} explicitly for the coupled 
two-dimensional quartic oscillator 
\be
V(x,y)=\frac{1}{2}(x^4+y^4)-\kappa\, x^2 y^2\,,
\label{vqo}
\ee
whose classical dynamics is almost chaotic\cite{btu,erda} in the 
limits $\kappa=1$ and $\kappa\to -\infty$, but in practice also for 
$\kappa=0.6$ (see, e.g., Ref.\cite{marta}).
\vspace*{-0.5cm}\\
We find an excellent agreement over the whole region. That the TF
kinetic-energy functional holds also for the oscillating exact
densities to a surprising degree has been noted long 
ago,\cite{brfest} but not understood until now. Similarly good 
numerical results are obtained also for the LVT \eq{lvt}, 
except very close to the classical turning points, for many
systems\cite{rbkm,br} with not too small particle numbers $N$.
\begin{figure}[h]
\begin{center}
\includegraphics[width=0.7\columnwidth,clip=true]{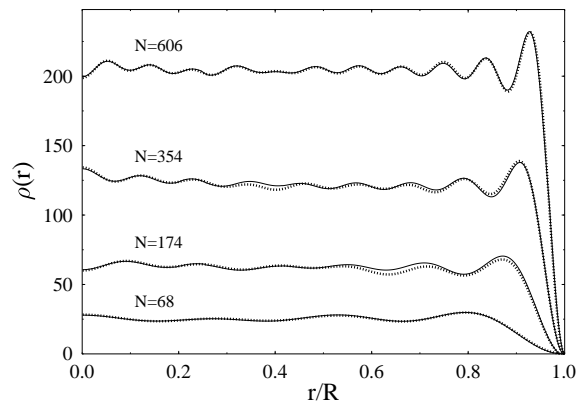}\vspace*{-0.3cm}
\caption[]{
Particle density in the 2-dimensional circular billiard 
with various particle numbers $N$ (units: $\hbar^2\!/2m=R=1$). 
Dotted lines: quantum results, solid lines: semiclassical results.\cite{br} 
\label{disk606}}
\end{center}
\end{figure}\vspace*{-0.5cm}

Next we present some results for the particle densities.
Figure \ref{disk606} shows $\rho(r)$ for four values of the 
number of $N$ particles bound in the two-dimensional circular 
billiard. The dotted line is the quantum result \eq{rho}, and the 
solid line the converged semiclassical result \eq{drhosc},
complemented by uniform approximations at the critical points
as explained in detail in Ref.\cite{br}\, Similar results are
obtained also for the kinetic-energy densities, and for other
types of potentials.\cite{rb,rbkm} 

It should be emphasized that, 
due to a factor $1/T_\gamma(\lambdab,\bfr)$ in the semiclassical
amplitudes ${\cal A}_\gamma(\lambdab,\bfr)$ in \eq{drhosc} - \eq{dtau1sc},
the sums over the orbits $\gamma$ converge much faster than in the 
trace formula \eq{trf} for the level density.

Note the {\it Friedel oscillations} in Fig.\ \ref{disk606} near 
the surface ($r=R$), which are characteristic of a fermionic system 
near a steep boundary. In our semiclassical theory, the Friedel 
oscillations are caused by the shortest orbit with one reflection 
from the boundary (in Refs.\cite{rb,rbkm} called the primitive 
``+'' orbit). Its regularized contribution to the particle density
of a spherical billiard in $D$ dimensions is\cite{rbkm}
\be
\delta\rho_{+}(r) = -\rho_{\text{TF}}^{(D)}\,2^\nu\Gamma(\nu+1)
                     \left(\frac{R}{r}\right)^{\!\nu-1/2}
                     \frac{J_\nu(z)}{z^\nu}\,,
\label{drfriedd} 
\ee
where $\rho_{\text{TF}}^{(D)}$ is the TF density, $\nu=D/2$,\, 
$z=2\,(R-r)\,p_\lambda/\hbar$, and $p_\lambda=(2m\lambdab)^{1/2}$ is
the smooth Fermi momentum. Integrating \eq{drfriedd} over the 
whole space, we obtain
\be
\delta N_+ = \int \d^D r\, \delta\rho_{+}(r)
   = - \frac{1}{2\pi^{D/2}\hbar^{D-1}}\,
       \frac{\Gamma(D/2)}{\Gamma(D)}\,p_\lambda^{D-1}\,{\cal S}_D\,,
\label{weylsurf}
\ee
where ${\cal S}_D = 2\pi^{D/2}R^{D-1}\!/\Gamma(D/2)$ is the 
hypersurface of the $D$-dimensional sphere. It is interesting
to note (see also Ref.\cite{zheng}) that \eq{weylsurf} corresponds precisely to the
{\it surface term} in the Weyl expansion\cite{weyl} of the
particle number $N(\lambdab)$ which varies smoothly with the
Fermi energy (the volume term being given by the TF theory).

\section{Inclusion of finite temperatures in the semiclassical theory}
\label{sectemp}

In the following we outline how to include finite temperatures in
the semiclassical formalism. Extensions of semiclassical trace
formulae to finite temperatures have been used long
ago in the context of nuclear physics\cite{mako} and more
recently in mesoscopic physics.\cite{ruj} We shall present 
here a derivation by means of a suitable folding function, 
which has proved useful also in the corresponding microscopic
theories\cite{bq} and allows for a straightforward generalization
to include other types of correlations.

For a grand-canonical ensemble of fermions embedded in a 
heat bath with fixed temperature, the variational energy is 
the so-called grand potential $\Omega$ defined by
\be
\Omega = \langle {\hat H} \rangle - TS - \lambda \langle {\hat N} \rangle\,,
\label{omega}
\ee
where ${\hat H}$ and ${\hat N}$ are the Hamilton and particle
number operators, respectively, $T$ is the temperature in energy 
units (i.e., we put the Boltzmann constant $k_{\text{B}}$ equal to unity),
$S$ is the entropy, and $\lambda$ the chemical potential.\footnote{
The quantities $S$ and $T$ without subscripts should not be confused 
with the actions $S_\gamma$ and periods $T_\gamma$ of the classical
orbits.} Note that both energy and particle number are conserved only
on the average. For non-interacting particles, we can write
the Helmholtz free energy $F$ as
\be
F = \langle {\hat H} \rangle - TS = 2\sum_n E_n \nu_n - TS\,.
\label{free}
\ee
Here $\nu_n$ are the Fermi occupation numbers
\begin{equation}
\nu_n = \frac{1}{1+\exp{\left(\frac{E_n - \lambda}{T}\right)}}\,,
\label{nuocc}
\end{equation}
and the entropy $S$ is given by
\begin{equation}
S=-2\sum_n \,[\nu_n \log\nu_n + (1-\nu_n) \log (1-\nu_n)] \,.      
\label{Sent}
\end{equation}
The chemical potential $\lambda$ is determined by fixing the
average particle number
\be
N = \langle {\hat N} \rangle = 2\sum_n \nu_n\, .                 
\label{avnum}
\ee
Note that all sums in \eq{free} -- \eq{avnum} and below run over the
complete (infinite) spectrum of the Hamiltonian ${\hat H}$.

It has been shown\cite{bq} that the above quantities $F$, $N$ 
and $S$ can be expressed in terms of a convoluted {\it 
finite-temperature level density} $g_T(E)$ defined by a 
convolution of the ``cold'' ($T=0$) density of states \eq{dos}
\begin{equation}
g_T(E):=\int_{-\infty}^\infty g(E')\,f_T(E-E') \, \d E'
       =\sum_n f_T(E-E_n)\,,                 
\label{gT}
\end{equation}
whereby the folding function $f_T(E)$ is given as
\be
f_T(E) = \frac{1}{4T\,\Cosh^2(E/2T)}\,.
\label{fT}
\ee
The free energy then is given by
\begin{equation}
F=2\int_{-\infty}^\lambda E\,g_T(E)\,\d E\,,
\label{gtrel}
\end{equation}
and the average particle number by
\be     
N = 2 \int_{-\infty}^\lambda g_T(E)\,\d E\,.
\ee
To show that the integral \eq{gtrel} gives the correct free energy 
\eq{free}, including the ``heat energy'' $-TS$, requires some 
algebraic manipulations. From $F$, the entropy $S$ can always be
gained by the canonical relation
\be
S=-\papa{F}{T}\,.               
\label{canonent}
\ee

The same convolution can now be applied also to the semiclassical
trace formula \eq{trf} for the oscillating part of the density
of states which we re-write as
\be
\delta g(E) \simeq \text{Re} \sum_{\text{PO}} {\cal A}_{\text{PO}}(E)\,
                   e^{i\Phi_{\text{PO}}(E)}
\label{trfc}
\ee
with the phase
\be
\Phi_{\rm PO}(E) = \frac{1}{\hbar}\,S_{\text{PO}}(E)-\frac{\pi}{2}\,\sigma_{\text{PO}}\,. 
\label{PhiE}
\ee
The oscillating part $\delta g_T(E)$ of the finite-temperature level density 
is obtained by the convolution of \eq{trfc} with the function $f_T(E)$
as in \eq{gT}. In the spirit of the stationary-phase approximation,
we take the slowly varying amplitude ${\cal A}_{\text{PO}}(E)$ outside of
the integration and approximate the action in the phase by
\be
S_{\text{PO}}(E') \simeq S_{\text{PO}}(E) + (E'-E)\,T_{\text{PO}}(E)\,,
\ee
so that the result becomes a modified trace formula
\be
\delta g_T(E) \simeq \text{Re} \sum_{\text{PO}} {\cal A}_{\text{PO}}(E)\,
                     {\tilde f}_T[\Tau_{\text{PO}}(E)]\,
                     e^{i\Phi_{\text{PO}}(E)},
\label{trft}
\ee
where 
\be
\Tau_{\text{PO}}(E) = T_{\text{PO}}(E)/\hbar\,
\ee
and the temperature modulation factor ${\tilde f}_T$ is given
by the Fourier transform of the convolution function $f_T$:
\be
{\tilde f}_T(\Tau) = \int_{-\infty}^\infty f_T(\omega)\,
                     e^{i\Tau\omega}\,{\rm d}\omega\,.
\ee
The Fourier transform of the function \eq{fT} is known\cite{batem} and yields
\be
{\tilde f}_T(\Tau) = \frac{\pi T\Tau}{\Sinh(\pi T\Tau)}\,. 
\label{modT}
\ee
The ``hot'' trace formulae \eq{trft} with the modulation factor
\eq{modT} has previously been obtained in Refs.\cite{mako,ruj}
The trace formula for the oscillating part of the free
energy then becomes\cite{book,mako} to leading order in $\hbar$
\be
\delta F \simeq \text{Re} \sum_{\text{PO}} {\cal A}_{\text{PO}}(\lambdab)\,
                \left(\frac{\hbar}{T_{\text{PO}}(\lambdab)}\right)^{\!2}    
                {\tilde f}_T[\Tau_{\text{PO}}(\lambdab)]\,
                e^{i\Phi_{\text{PO}}(\lambdab)}.
\label{delF}
\ee

For the spatial densities we can proceed exactly in the same way. 
For the particle density, e.g., the microscopic expression \eq{rho}
is replaced by
\be
\rho_T(\bfr) = 2 \sum_n |\phi_n(\bfr)|^2\nu_n\,,
\label{rhoT}
\ee
where the sum again runs over the complete spectrum. Starting 
from the semiclassical expression \eq{drhosc} for $\delta\rho(r)$
at $T=0$, we rewrite it as
\be
\delta\rho_0(\lambdab,\bfr) \simeq \text{Re} \sum_\gamma {\cal A}_\gamma(\lambdab,\bfr)\,
                                   e^{i\Phi_\gamma(\lambdab,\bfr)},
\label{drhoscfoll}
\ee
where $\Phi_\gamma(\lambdab,\bfr)$ is the phase \eq{phaser}. 
The finite-$T$ expression is given by the convolution integral
\be
\delta\rho_T(\lambdab,\bfr) \simeq \int_{-\infty}^{\lambdab} \delta\rho_0(\lambdab-E,\bfr) 
                                   f_T(E)\, \d E\,.
\ee
Expanding the phase under the integral as above, we arrive at
\be
\delta\rho_T(\lambdab,\bfr) \simeq \text{Re} \sum_\gamma {\cal A}_\gamma(\lambdab,\bfr)\,
                            {\tilde f}_T[\Tau_\gamma(\lambdab,\bfr)]\,e^{i\Phi_\gamma(\lambdab,\bfr)}\,,
\label{drhoscT}
\ee
where $\Tau_\gamma(\lambdab,\bfr)=T_\gamma(\lambdab,\bfr)/\hbar$. The corresponding 
expressions for the temperature-dependent kinetic-energy densities 
are obvious.

For the smooth parts of the densities, we recall that the ETF theory 
at $T>0$ is well known (see, e.g.\ Ref.\cite{etft}, where expressions
up to 4-th order in $\hbar$ are given, and the literature quoted therein).

Other types of correlations can be included in the semiclassical
theory in the same way, as soon as a suitable folding function 
$f_{\text{corr}}(E)$ -- corresponding to $f_T(E)$ in \eq{fT} -- and its Fourier 
transform are known. One example is given by the pairing correlations 
discussed in the following section.

\section{Inclusion of pairing correlations in the BCS approximation}
\label{secbcs}

A self-consistent microscopic approach to include pairing correlations
is given by the Hartree-Fock-Bogolyubov (HFB) approach; we refer to 
an extended article\cite{bq} 
for a recapitulation of this theory and the relevant literature. In the
simplified BCS approach with constant paring gap $\Delta$, the total
energy of a system is written as
\be
E_{\text{BCS}} = \sum_n E_n v_n^2 - \Delta \sum_n u_n v_n\,,
\label{ebcs}
\ee
where the sum goes over the complete spectrum (including all
degeneracies) and the occupation numbers $u_n$ and $v_n$ are given by
\bea
v_n & = & \frac{1}{\sqrt{2}}\left[1+\frac{(\lambda-E_n)}{{\cal E}_n}\right]^{1/2},\nonumber\\
u_n & = & \sqrt{1-v_n^2}\,.
\label{vnun}
\eea
Hereby ${\cal E}_n$ is the so-called quasiparticle energy
\be
{\cal E}_n(\lambda) = {_+\!}\sqrt{(\lambda-E_n)^2+\Delta^2}\,.
\label{eqp}
\ee
It was shown\cite{bq} that the BCS energy \eq{ebcs} is correctly
given, including the pair condensation energy
\be
E_p = -\,\Delta\sum_n u_n v_n\,,
\label{Epair}
\ee
by the convolution integral
\be
E_{\text{BCS}} = \int_{-\infty}^\lambda E\,f_\Delta(E)\,\d E\,,
\label{ebscfol}
\ee
where the folding function $f_\Delta(E)$ is defined as
\be
f_{\Delta}(E) := \frac{\Delta^2}{2\left[E^2+\Delta^2\right]^{3/2}}.
\label{fDelta}
\ee
The Fermi energy $\lambda$ in all above expressions is fixed by the 
average particle number:
\be
N = \int_{-\infty}^\lambda g_\Delta(E)\,\d E = \sum_n v_n^2\,.
\ee
The ``paired'' level density $g_\Delta(E)$ is given by
\be
g_\Delta(E) = \sum_n \frac{\Delta^2}{2\left[{\cal E}_n(E)\right]^3}.
\label{gDelta}
\ee
The Fourier transform of $f_\Delta(E)$ is found\cite{batem} to be
\be
{\widetilde f}_\Delta(\Tau) = \Delta\Tau K_1(\Delta\Tau)\,,
\ee
where $K_1(z)$ is a modified Bessel function.\cite{abro} Hence, 
replacing ${\widetilde f}_T$ in \eq{trft} by ${\widetilde f}_\Delta$,
the semiclassical trace formula for the oscillating part of the
paired level density becomes
\be
\delta g_\Delta(E)  \simeq \text{Re} \sum_{\text{PO}} {\cal A}_{\text{PO}}(E)\,
                           {\tilde f}_\Delta[\Tau_{\text{PO}}(E)]\,
                           e^{i\Phi_{\text{PO}}(E)}.
\label{trfDelta}
\ee
The trace formula for the oscillating part of the total BCS energy
becomes, analogously to \eq{delF},
\be
\delta E_{\rm BCS} \simeq \text{Re} \sum_{\text{PO}} {\cal A}_{\text{PO}}(\lambdab)\,
                   \left(\frac{\hbar}{T_{\text{PO}}(\lambdab)}\right)^{\!2}    
                   {\tilde f}_\Delta[\Tau_{\text{PO}}(\lambdab)]\,
                   e^{i\Phi_{\text{PO}}(\lambdab)}.
\label{delEBCS}
\ee
That for the pair condensation energy, using $E_p=\Delta\, 
\partial E_{\rm BCS}/\partial\Delta$ and exploiting a recurrence relation
for the Bessel functions,\cite{abro} becomes
\be
\delta E_p \simeq  \Delta^2\, \text{Re} \sum_{\text{PO}} {\cal A}_{\text{PO}}(\lambdab)\,
                   K_0[\Tau_{\text{PO}}(\lambdab)]\,
                   e^{i\Phi_{\text{PO}}(\lambdab)}.
\label{delEpair}
\ee
A similar result has recently been obtained in Ref.\cite{olof}

For the spatial densities we can, in principle, proceed as above.
The pair-correlated particle density is quantum-mechanically given by\cite{bq}
\be
\rho_\Delta(\bfr) = \sum_n |\phi_n(\bfr)|^2 v_n^2\,.
\ee
The semiclassical expression of its oscillating part becomes, similarly
as above,
\be
\delta\rho_\Delta(\lambdab,\bfr) \simeq \text{Re} \sum_\gamma {\cal A}_\gamma(\lambdab,\bfr)\,
                           {\tilde f}_\Delta[\Tau_\gamma(\lambdab)]\,e^{i\Phi_\gamma(\lambdab,\bfr)}\,.
\label{drhoscDelta}
\ee
Corresponding results hold for the pair-correlated kinetic-energy densities.

This is, however, not the end of the story. If one wants to express the
pair-condensation energy \eq{Epair} as a space integral, one requires 
an anomalous density matrix $\kappa(\bfr,\bfr')$, defined by\cite{bq}
\be
\kappa(\bfr,\bfr') = \sum_n \phi_n(\bfr)\phi_{\bar n}(\bfr')\,,
\ee
where ${\bar n}$ refers to the time-reversed state of $n$. The
semiclassical evaluation of this anomalous density matrix is the
object of our ongoing research.

\newpage


\end{document}